\documentclass[pra,twocolumn,showpacs,preprintnumbers,amsmath,amssymb,floatfix,superscriptaddress,nofootinbib]{revtex4}

\usepackage{graphicx}
\usepackage{epstopdf}
\usepackage{dcolumn}
\usepackage{bm}

\begin{document}

\title{
Possibility of ``magic'' co-trapping of two atomic species in optical lattices
}
\author{Muir J. Morrison}
\affiliation{
Department of Physics, University of Nevada, Reno, Nevada 89557}

\author{V. A. Dzuba}
\affiliation{School of Physics, University of New South Wales, Sydney 2052,
Australia}
\affiliation {
Department of Physics, University of Nevada, Reno, Nevada 89557}

\author{A. Derevianko}
\affiliation {
Department of Physics, University of Nevada, Reno, Nevada 89557}

\date{\today}

\begin{abstract}
Much effort has been devoted to removing differential Stark shifts for atoms trapped in specially tailored ``magic'' optical lattices, but thus far work has focused on a single trapped atomic species. In this work, we extend these ideas to include two atomic species sharing the same optical lattice. We show qualitatively that, in particular, scalar $J=0$ divalent atoms paired with non-scalar state atoms have the necessary characteristics to achieve such Stark shift cancellation. We then present numerical results on ``magic" trapping conditions for $^{27}$Al paired with $^{87}$Sr, as well as several other divalent atoms.
\end{abstract}


\maketitle

\section{Introduction}

Cold atoms and molecules have proven to be a useful platform for a variety of precision measurement and quantum information experiments. Optical lattices provide a method of trapping cold atoms and molecules that minimizes Doppler shifts and allows long measurement times. However, the trapping field causes shifts in the internal energy levels via the ac Stark effect, and these shifts depend on the intensity of the trapping lasers. This problem has been considered in detail in the context of
optical lattice clocks (see, e.g., ~\cite{KatTakPal03,YeKimKat08,RosGheDzu09,BelDerDzu08Clock}). The solution employed in these works is to use an optical lattice operating at ``magic"  trapping conditions (e.g., magic wavelength)  for which the energy levels of interest experience the same shift, even as the atom or molecule moves within the lattice. This leaves the transition frequency of interest unchanged, as long as higher-order effects are negligible. In another recent work, the problem of removing Stark and Zeeman shifts simultaneously was considered for alkali-metals~\cite{Der10DoublyMagic}. In each of these works, cancellation of the Stark shift was achieved for a specific clock or qubit transition in a single atomic species.

In other experiments, simultaneous trapping of two atomic species has been achieved using double magneto-optical traps (MOTs)~\cite{ModFerRoa01,HadStaDie02,MudKraSin02}, combined with an optical dipole trap in~\cite{MudKraSin02}, with the primary goal of sympathetically cooling atomic species that are difficult to cool directly. However, in these experiments the internal states of the atoms are heavily perturbed by the MOT, which is disadvantageous for precision measurements.
Comagnetometry experiments also utilize two atomic species sharing the same volume. One such apparatus was used to create a highly accurate nuclear spin gyroscope~\cite{KorGhoRom05}, while other groups have used comagnetometers in searches for physics beyond the standard model, such as setting upper bounds on P,T-violating permanent electric dipole moment~\cite{RosChu00} as well as bounds on Lorentz and {\em CPT} violation~\cite{CanBeaPhi04}.

In the present work, we aim to merge these ideas by considering two atomic species trapped in the same ``magic" optical lattice. Our goal is to achieve Stark-shift cancellation simultaneously in both atomic species, for a single two-level transition in each. We first introduce the equations underlying the ac Stark effect. These equations guide our selection of favorable atom pairs to analyze in detail. Finally, we present results of numerical calculations on $^{27}$Al paired with Sr, Yb, and other divalent atoms.

\section{Stark shift}

\begin{figure}[h]
   \begin{center}
   \includegraphics*[scale=0.45]{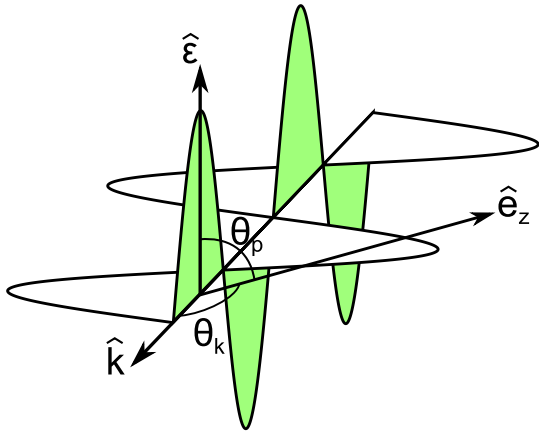}
   \end{center}
   \caption{(Color online)
   Relation of angles to unit vectors for linearly polarized light. $\hat{k}$, $\hat{\varepsilon}$, and $\hat{e}_z$ are the lattice wavevector, lattice polarization, and quantization axis, respectively. For circular polarization, we take $\hat{\varepsilon}$ to be complex, following the Jones calculus conventions. In this case $\theta_k$ is unchanged, while $\theta_p$ is no longer well-defined.
    \label{Fig:geom}
    }
   \end{figure}

The derivation of the Stark shift follows closely the treatment in~\cite{RosGheDzu09} and~\cite{FlaDzuDer08}, so only a summary of the results will be given here. The ac Stark shift of an atomic state $| F, M_F\rangle$ with total angular momentum $F$ and projection $M_F$ can be written as~\footnote{We use atomic units except where otherwise noted.}
\begin{equation}
\delta E_{FM_F}(\omega)= - \alpha_{FM_F}^{tot}(\omega) \left(\frac{\mathcal{E}_L}{2}\right)^{2} \, ,
\label{Eq:dE}
\end{equation}
where $\mathcal{E}_L$ and $\omega$ are the amplitude and frequency of the lattice light field. The frequency-dependent atomic-structure pre-factor $\alpha_{FM_F}^{tot}(\omega)$ is the total ac polarizability of the state; it may be decomposed in terms of irreducible tensor operators as
\begin{widetext}
  \begin{equation}
      \alpha_{FM_F}^{tot}(\omega)= \alpha_F^S (\omega)
    + (\hat{k}\cdot\hat{e}_z) \, \mathcal{A} \, \frac{M_F}{2F}\alpha_F^a (\omega)
       + \frac{1}{2}\left(3\left|\hat{\varepsilon}\cdot\hat{e}_z\right|^{2}-1\right)
        \frac{3{M_F}^{2}-F(F+1)}{F(2F-1)}\alpha_F^T (\omega) \label{Eq:alphaTot} \, .
  \end{equation}
\end{widetext}
 Here $\mathcal{A}$ is the degree of circular polarization, $\alpha_F^S$, $\alpha_F^a$, and $\alpha_F^T$ are the scalar, vector, and tensor polarizabilities, respectively, and the unit vectors are the lattice laser wavevector ($\hat{k}$), polarization ($\hat{\varepsilon}$), and quantization axis ($\hat{e}_z$). Fig.~\ref{Fig:geom} depicts the geometry.
The quantization axis is defined by an applied static magnetic field, whose orientation relative to the optical lattice may be arbitrary. This ``quantizing magnetic field" ensures that $M_F$ remains a ``good" quantum number for the ac Stark-effect perturbation formalism, from which Eq.~(\ref{Eq:alphaTot}) is derived; the energy shifts due to the Stark effect must be small compared with the Zeeman splitting of the magnetic sublevels.
It is important to note that the total polarizability depends not only on the spatial orientation of $\hat{k}$, $\hat{\varepsilon}$, and $\hat{e}_z$, but also on the lattice frequency $\omega$ as well.
Also useful is the total differential polarizability $\Delta\alpha$ between two states $|F^{\prime},M_F^{\prime}\rangle$ and $|F,M_F\rangle$, given simply by
\begin{equation}
  \Delta\alpha(\omega) = \alpha_{F^{\prime}M^{\prime}_F}^{tot}(\omega) - \alpha_{FM_F}^{tot}(\omega) \, .
  \label{Eq:Dalpha}
\end{equation}
Analogously, we will use the scalar differential polarizability, defined as
\begin{equation}
  \Delta\alpha^{S}(\omega) = \alpha_{F^{\prime}M^{\prime}_F}^{S}(\omega) - \alpha_{FM_F}^{S}(\omega)
  \label{Eq:Dalpha}
\end{equation}
and also the vector and tensor differential polarizabilities, defined similarly.

For linearly polarized light, the vector contribution vanishes ($\mathcal{A}=0$), and $\hat{\varepsilon}\cdot\hat{e}_z = \cos{\theta_p}$, where $\theta_p$ is the angle between the polarization and quantization unit vectors. Then the total polarizability becomes
\begin{widetext}
\begin{equation}
\alpha_{FM_F}(\omega)= \alpha_F^S (\omega)
+ \frac{1}{2}\left(3\cos^{2}{\theta_p}-1\right)\frac{3{M_F}^{2}-F(F+1)}{F(2F-1)}\alpha_F^T (\omega) \, .
\label{Eq:alphaLin}
\end{equation}
\end{widetext}
In the case of circularly polarized light, $\hat{k}\cdot\hat{e}_z = \cos{\theta_k}$ and $\left|\hat{\varepsilon}\cdot\hat{e}_z\right|^{2} = \frac{1}{2}\sin^{2}{\theta_k}$, where $\theta_k$ is the angle between the wavevector and quantization unit vectors. $\hat{\varepsilon}$ is taken to be complex and constant in time, rather than real and time varying. This means that, although counterintuitive, $\left|\hat{\varepsilon}\cdot\hat{e}_z\right|^{2}$ is time-independent even for circular polarization. The total polarizability is then
\begin{widetext}
\begin{equation}
\alpha_{FM_F}(\omega)= \alpha_F^S (\omega)
+ \mathcal{A}(\cos{\theta_k})\frac{M_F}{2F}\alpha_F^a (\omega)
 + \frac{1}{2}\left(\frac{3}{2}\sin^2{\theta_k}-1\right)\frac{3{M_F}^{2}-F(F+1)}{F(2F-1)}\alpha_F^T (\omega) \, .
\label{Eq:alphaCirc}
\end{equation}
\end{widetext}
It should be noted that in both cases, the Stark shift depends on only one angle (either $\theta_p$ or $\theta_k$, but not both). This means that for purely circularly or linearly polarized light, we have only two ``handles," not three, with which to achieve ``magic" conditions: a single angle and the lattice frequency $\omega$.

It should be noted that the tabulated formulae are valid only for a sufficiently weak magnetic field.
For finite B fields, the Stark shift acquires an additional cross-term formed as a product of the Zeeman and vector light shifts; this provides an additional knob for tuning magic conditions (see~\cite{LunSchPor10,Der10Bmagic}); we will comment on this in Section~\ref{Sec:RbCs}.

\section{Selecting Atoms for Trapping}
\label{chooseAtoms}
Guided by the above equations, we state some general guidelines in selecting two atomic species to trap. Consider first the vector and tensor polarizabilities. These are rank 1 and rank 2 irreducible tensor operators, respectively. Triangular selection rules dictate that $\alpha_F^T$ must vanish for $F=1/2$ states, and that both $\alpha_F^T$ and $\alpha_F^a$ must vanish for $F=0$ states. Alternately, the polarizabilities of $F=0$ states have no dependence on the angles $\theta_p$ or $\theta_k$.

This leads us to consider divalent atoms, such as Sr or Yb, which have $^1\!S_0$ ground states. The bosonic isotopes have nuclear spin $I=0$, so that $F=J=0$: due to the angular selection rules $\alpha_F^T$ and $\alpha_F^a$ must vanish. The fermionic isotopes have $I\ne0$, and so (for $I>1/2$), $\alpha_F^T$ and $\alpha_F^a$ do not vanish identically. However, $\alpha_F^T$ and $\alpha_F^a$ are nonzero only due to a highly suppressed contribution from the hyperfine interaction (HFI), and so $\alpha_F^S$ dominates. For optical transitions between $J=0$ electronic states, the differential polarizability is almost entirely due to the scalar polarizability. Therefore the light shift, being of scalar nature, does not depend on atomic spatial orientation.
 We conclude that for $0-0$ clock transitions in divalent atoms, the dependence on geometry is negligible and ``magic" conditions are set entirely by the lattice frequency.
If one of our two trapped atomic species is a scalar atom, we must use a ``magic" lattice frequency for that atom. Since the frequency becomes fixed, we have to achieve magic conditions for the second species solely by manipulating one angle, $\theta_p$ or $\theta_k$ (for fixed circular or linear polarization).

While the discussion so far has been of a general nature, for illustration let us consider microwave hyperfine transitions attached to a $J=1/2$ electronic state  in our second atomic species. The relevant atoms may include the ground state alkali-metal atoms ($s_{1/2}$ ground states) or group III atoms ($p_{1/2}$ ground states).
To minimize Zeeman shifts, we will work with $M_F=0$ states exclusively, which also eliminates the vector term in Eq.~(\ref{Eq:alphaCirc}).
There are some technical but important peculiarities associated with computing Stark shifts between levels of the same hyperfine manifold: we are considering the Stark shift of hyperfine levels attached to the same electronic
state. To the leading order, the shift is determined by the properties of the underlying electronic state. However, because the electronic state for
both hyperfine levels is the same, the scalar Stark shift of both levels is the same.
An apparent difference between the two
clock levels is caused by the hyperfine interaction (HFI), and the rigorous analysis involves
so-called HFI-mediated polarizabilities~\cite{RosGheDzu09}. Similarly, when neglecting the HFI, the  {\em tensor} shift for the $J=1/2$ state
vanishes due to the angular selection rules; we are also led to a necessity to treat the tensor shift using the HFI-mediated
polarizabilities.

As demonstrated in Ref.~\cite{BelDerDzu08Clock}, for $M_F=0$ clock or qubit states and fixed $\mathbf{B} \| \hat{k}$ or $\mathbf{B}\perp \hat{k}$ geometry neither Rb nor Cs can be magically trapped. However, the magic trapping can be achieved for Al and Ga. Our consideration here is more general, as we also consider varying rotation angles. Still even in our more general setup we do not find magic angles for the $M_F=0$ clock or qubit states in Cs and Rb (again this statement is limited only to the weak B field regime). We focus below on Al, where the required magic angles can be found.

To summarize, we choose a scalar divalent atom and a non-scalar valence state atom to trap simultaneously. We choose for our optical lattice a ``magic" frequency for the divalent atom, so the differential polarizability for its optical transition of interest is zero. At this lattice frequency, the differential polarizability for the microwave transition in the non-scalar atom is non-vanishing, in general. But by tuning $\theta_p$ or $\theta_k$ (depending on lattice polarization), we aim to create ``magic" conditions in the non-scalar atom as well.


We perform numerical calculations to find ``magic" conditions for particular pairs of atoms. Utilizing a variety of relativistic many-body techniques, as described in~\cite{BelSafDer06}, we calculate the atomic polarizabilities in Eq.~(\ref{Eq:alphaLin}) and Eq.~(\ref{Eq:alphaCirc}). We then numerically search for values of $\theta_{k}$ or $\theta_{p}$ that will make $\Delta\alpha$ vanish in the non-scalar atom. We focus our attention on $^{87}$Sr paired with $^{27}$Al, although we also present results for $^{27}$Al paired with other divalent atoms. In $^{87}$Sr, the transition of interest is between the $5s^2$~$^1S_0$ ground state and the $5s5p$~$^3P_0$ metastable state~\cite{TakHonHig05}. In $^{27}$Al, the transition of interest is between the $| F=2, M_F=0\rangle$ and $| F=3, M_F=0\rangle$ hyperfine states, both in the $3p_{1/2}$ electronic ground state. We work with $M_{F}=0$ states exclusively in Al so that (to first order) Zeeman shifts vanish.

\section{Results}
\begin{figure} [h]
\begin{center}
\includegraphics*[scale=0.9]{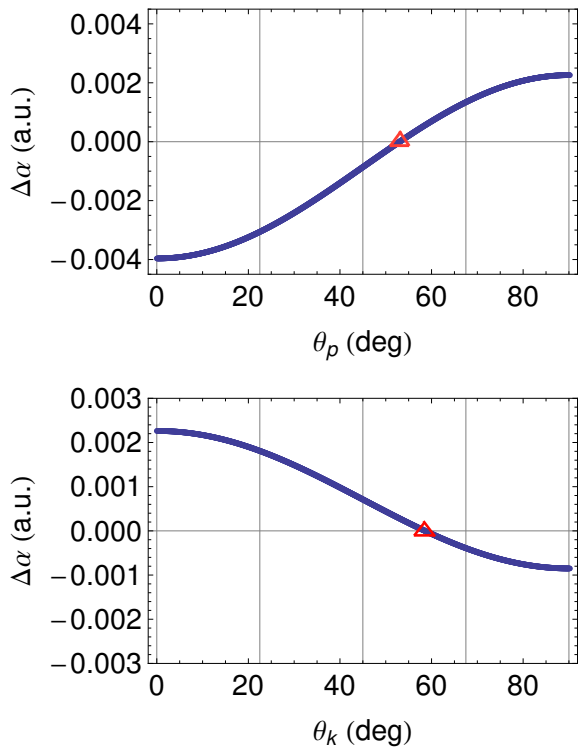}
\end{center}
\caption{(Color online)
Differential polarizability (Eq.~\ref{Eq:Dalpha}) between $| F=3, M_F=0\rangle$ and $| F=2,M_F=0\rangle$ states in $^{27}$Al as a function of angle. The top graph is for a linearly polarized lattice, while the bottom graph is for a circularly polarized lattice. Note $\theta_p$ is the angle between the polarization and quantization axis, while $\theta_k$ is the angle between the wavevector and quantization axis, as shown in Fig.~\ref{Fig:geom}. The magic angle is the value of $\theta$ for which $\Delta\alpha = 0$, highlighted with red triangles. This is for a lattice at 813.42~nm, the ``magic" wavelength for the optical clock transition in $^{87}$Sr.
 \label{Fig:Al_813}
 }
\end{figure}

In Fig.~\ref{Fig:Al_813} we plot results representative of our numerical calculations. We graph the differential polarizability between the two hyperfine states in $^{27}$Al as a function of angle, for linear and circular lattice polarizations.

While section~\ref{chooseAtoms} suggests that Stark-shift cancellation may be possible for $^{27}$Al at arbitrary lattice frequencies, it is not obvious if ``magic" angles for $^{27}$Al actually exist at the ``magic" wavelengths for the divalent atoms of interest. Considering Eq.~(\ref{Eq:alphaLin}) and Eq.~(\ref{Eq:alphaCirc}), we note that the prefactor on the tensor term varies from approximately $1$ to $-\frac{1}{2}$ for linear polarization and from $\frac{1}{4}$ to $-\frac{1}{2}$ for circular polarization. This means it is highly probable a ``magic" angle exists if $\alpha_{F^{\prime}}^T$ and $\alpha_F^T$ are at least a few times greater than $\alpha_{F^{\prime}}^S$ and $\alpha_F^S$ (barring a happenstance cancellation of $\alpha_{F^{\prime}}^T$ and $\alpha_F^T$). Our calculations reveal that, over the range of wavelengths studied (see Table~\ref{Table:ThetaList}), $\alpha_{F^{\prime}}^T$ and $\alpha_F^T$ vary from 2 orders of magnitude larger than $\alpha_{F^{\prime}}^S$ and $\alpha_F^S$ to roughly the same order of magnitude. Only near the Al resonance at 394~nm does the scalar polarizability dominate. This suggests that ``magic" angles should exist for many pairings of Al with scalar divalent atoms.

Table~\ref{Table:ThetaList} shows results of our numerical calculations, confirming the above assertions. We tabulate the lattice frequency for which the Stark shift vanishes in each divalent species, and ``magic" angles for which the Stark shift in $^{27}$Al also vanishes for linearly and circularly polarized lattices. The angles $\theta_{p}$ and $\theta_{k}$ are as depicted in Fig.~\ref{Fig:geom} and defined in Eqs.~(\ref{Eq:alphaLin}) and~(\ref{Eq:alphaCirc}), respectively.
\begin{table}[h]
\caption{ Values of $^{27}$Al ``magic'' angles at ``magic'' wavelengths for divalent atoms. Values for $\lambda_{magic}$ from Ref.~\cite{DerObrDzu09}.
 \label{Table:ThetaList}}
\begin{ruledtabular}
\begin{tabular}{cccc}
  &  $\lambda_{magic} [nm]$   &    $\theta_p$ (linear pol.)   &   $\theta_k$ (circ. pol.) \\
\hline
Hg  &  362  &  58.89$^{\circ}$    &   46.95$^{\circ}$  \\
Zn  &  416  &  48.10$^{\circ}$    &   70.82$^{\circ}$  \\
Cd  &  419  &  48.82$^{\circ}$    &   68.62$^{\circ}$  \\
Mg  &  466  &  51.91$^{\circ}$    &   60.73$^{\circ}$  \\
Ca  &  739  &  52.90$^{\circ}$    &   58.54$^{\circ}$  \\
Yb  &  759  &  52.91$^{\circ}$    &   58.52$^{\circ}$  \\
Sr  &   813  &  52.93$^{\circ}$    &   58.47$^{\circ}$  \\
\end{tabular}
\end{ruledtabular}
\end{table}
%

An important consideration is the sensitivity of transition frequency to experimental error in the magic angle. We can write the fractional clock shift as
\begin{equation}
\frac{\Delta\nu}{\nu} = - \frac{I_L}{c\nu}\Delta\alpha
\label{Eq:dnu/nu}
\end{equation}
where $I_L$ is the laser intensity and $\Delta\alpha$ denotes the differential polarizability between the upper and lower clock levels, as defined in Eq.~(\ref{Eq:Dalpha}). For Al in a lattice with Sr ($\lambda = 813$~nm), a misalignment of 0.1$^\circ$ corresponds to $\Delta\alpha \approx 10^{-5}$~a.u. If we assume a lattice intensity on the order of 10~kW/cm$^2$, we are lead to a fractional clock shift of the microwave transition $\frac{\delta\nu}{\nu}\approx10^{-15}$. This effect would be somewhat worse for shorter wavelength lattices, since the polarizabilities tend to increase faster than $\nu$.

\section{Rubidium and Cesium}
\label{Sec:RbCs}
So far we focused our attention on the weak-field regime to reduce Zeeman-sensitivity of the  $M_F' - M_F=0$ transition.
If we relax the weak-field requirement, magic conditions for such transitions can be found by varying the strength of the B-field. Magic trapping conditions then exist for all trapping wavelengths (barring atomic resonances), see Fig. 2 of Ref.~\cite{Der10Bmagic}.  The required B field strengths are of a few Gauss. For example, one could read off Fig. 2 of Ref.~\cite{Der10Bmagic} that at the magic 813 nm wavelength for Yb, trapping of $^{133}$Cs is magic for $B=2.9 G$.

In another  approach one may choose to operate on multiphoton $M_F \rightarrow - M_{F}$ transitions\cite{Der10DoublyMagic}; such transitions have ``magic'' values of $B$ fields, where the Zeeman sensitivity is removed at points where $d\nu/dB=0$. Again there is a range of lattice wavelengths where  the trapping may be made magic by varying
the angle between the lattice wavevector and the quantizing B-field. For  example, for $^{133}$Cs $|4,3\rangle \to  |3,-3\rangle$ transition,
the acceptable ranges of magic wavelengths in nm are 898--1591, 863--880, and 512--796 (see Table I of Ref.\cite{Der10DoublyMagic}). Since the magic wavelengths of Ca and Yb fall within these ranges, these species can be co-trapped with Cs.

\section{Conclusion}
We have presented a method to simultaneously remove Stark shifts for two atomic species trapped in a shared optical lattice. One transition in each atom may be made insensitive to the intensity of the trapping laser fields.
This method could prove useful for quantum information and precision measurement experiments. For example, two different atomic species may be used for executing quantum multi-particle gates and state-dependent transport
(see, e.g.~\cite{WeiBelDer10}). Operating the common lattice at the simultaneous magic trapping conditions would remove detrimental Stark-induced decoherences for both types of qubits.


\end{document}